\documentclass[a4paper,12pt]{article}

\begin{document}
\date{\today}

\title{Simple operator formulation of the Bakamjian-Thomas approach to heavy quark current , with generalisation to HQET, and with applications to transitions of $\Lambda_b$.
}

\author{ D.~Becirevic$^a$, A.~Le~Yaouanc$^a$, V. Mor\'enas $^b$, L.~Oliver$^a$}
\maketitle

{\center $^a$~Laboratoire de Physique Th\'eorique (Unit\'e
Mixte de Recherche 8627 du CNRS),
Universit\'e de Paris-Sud et Universit\'e de Paris-Saclay, B\^at. 210, Centre d'Orsay, 91405 Orsay-Cedex, France. $^b$ Laboratoire de Physique de Clermont (Unit\'e Mixte de Recherche 6533 CNRS/IN2P3), Universit\'e Blaise Pascal, 4 Avenue Blaise Pascal, TSA 60026, 63171 Aubi\`ere Cedex, France \footnote{Preprint  LPT-Orsay-19-27 Orsay and preprint Clermont}}


\begin{abstract}
Starting from the general formulation of current matrix elements in the Bakamjian-Thomas (BT) approach to quark models, certain aspects of their heavy quark limit are exhibited in an intuitive form, allowing a common treatment for any number of quarks and angular momenta $j,j'$. We use first the general Falk interpretation of Isgur-Wise functions as overlaps of light parton states in motion. Then, in BT, for small collinear velocities, the motion of these states is described by a very simple dimensionless differential operator ${\cal O}$ acting on the internal wave functions of the light quarks, which is a relativistic generalisation of the dipole operator times the quark mass : it appears as the generator of Lorentz transformations for $N$ free light particles. It has a simple $O(3)_{space} \times SO(3)_{spin}$ structure $l=1 ;~s=0,1 ;~j=1$, restricting automatically the possible transitions. 
The slopes of the Isgur-Wise functions at $w=1$ for $l=0 \to l=0$ or $l=0 \to l=1$ are matrix elements of ${\cal O}^2$ or ${\cal O}$ respectively, leading to a straightforward algebraic demonstration of the Bjorken sum rule. 

An expression of the Isgur-Wise functions, manifestly covariant for collinear Lorentz transformations, is then given in terms of ${\cal O}$, by exponentiation, leading also to simple expressions of the higher derivatives. The formulation has immediate generalisation to HQET by simply substituting for ${\cal O}$ the generator of the collinear Lorentz transformation . Whence a compact formulation of HQET sum rules for higher derivatives(e.g. curvature for $\Lambda_b$ baryons), as well as selection rules, implying automatically their validity in BT quark models.
\end{abstract}

\section{Introduction}
Since a number of years, we have extensively developped for heavy mesons the BT (Bakamjian-Thomas) manifestly covariant approach initiated in Covariant quark model in the heavy mass limit of form factors \cite{covariant} (see for instance \cite{quantitative}). We are now preparing a paper for baryons(\cite{xxx}). In this approach, starting from the initial formulation with two dimensional description of quark spins, we were able, by using a 4-dimensional, covariant, spin description, to put it into a form manifestly covariant in the heavy quark limit, and which allows then  straightforward calculation of Isgur-Wise functions. Since we were also able to demonstrate in several practically important cases
the validity of certain HQET sum rules of the Bjorken-Uraltsev type for mesons in the same approach~(see refs \cite{exact,quantitative,nousUraltsev,sigmaMBT}), a rather satisfactory model for heavy hadrons is provided\footnote{Though, it is less satisfactory as soon as one tries to apply it to the order ${\cal O}(1/m_Q)$ where the model is not covariant anyway ; the sum rules like the one of Voloshin, implying the energies of the states are not satisfied.}. 

It is true, however, that the procedure is becoming increasingly complex when one considers several light quarks, as for baryons. In the present paper, while sticking to the same BT general approach, we are following a different method : we stick to the general formalism in which we have first defined the BT approach to matrix elements of currents \footnote{
This formalism, from which the manifestly covariant one is derived for particular cases, is presented in the beginning of the paper  \cite{covariant}}. This general formalism uses the two dimensional description of spin with Wigner rotations, and while it is not manifestly covariant, it has the advantage to treat any spin and any number of light quarks in a homogeneous way, and has a transparent interpretation by clear separation of the Lorentz transformations of space and spin, and the use of the standard three-dimensional angular momentum analysis. Thanks to this, we aim at presenting certain simple and more general results, valid for heavy mesons, baryons and possibly other multiquarks. No need to say that the two formalisms are strictly equivalent since the manifestly covariant formulae are in fact derived in each case from the 2-dimensional one. A sketch of this alternative method was already given in the paper  "Exact duality and Bjorken sum rule in heavy quark models \`a la Bakamjian-Thomas"
 ref. (\cite{exact}) to demonstrate the Bjorken sum rule.

\subsection{Falk factorisation : "overlaps" as the main dynamical concept}
A first important step is that the full matrix element of heavy quark currents in its original form can be factorised, in the heavy quark limit, into  : 1) the scalar product or "overlap" of states in motion, representing the light quarks in the field of the heavy source, times 2) the trivial factor of the active heavy quark current, with suitable CG coefficients to combine the spin
of the heavy quark and the $j$ of the light system \cite{exact}. This is the Falk factorisation \cite{falk} for heavy quark currents. It can be written as 
\begin{eqnarray} \label{falkPost}
<J'J_z'v'|{\cal J}|J J_z v>=
C_{\frac {1}{2} j'}(J'J_z';s_1',j_z' ) C_{\frac {1}{2}~j}(JJ_z;s_1,j_z ) \nonumber \\
\times<1/2, s_1'v'|{\cal J}|1/2 s_1 v>~<j'j_z' v'~|~j j_z v>
\end{eqnarray},
where $<1/2, s_1'|{\cal J}|1/2, s_1>$ is the free heavy quark current matrix element, or more explicitly, in Dirac spinor notation :
\begin{equation} 
<1/2 ,s_{1}'v'|{\cal J}|1/2, s_{1}v>=\bar{u}_{s_{1}'}(v') \Gamma u_{s_{1}}(v)
\end{equation}
$\Gamma$ being a generic Dirac matrix. On the other hand, the overlap $<j'j_z'v'~|~j j_z v>$ refers to the light cloud component and $j,j'$ are the spins of the light component states.

We leave aside the CG coefficients and the heavy current factors, and from now on, the term "overlap" will designate the light cloud part, which is the one containing dynamics. This can be taken as a new, simpler definition of the Isgur-Wise functions \footnote{The general definition by Falk overlaps was already used and discussed in our Lorentz papers \cite{LorentzI, LorentzII} ; a difference however is that angular momenta $j$ were described by covariant polarisation tensors or spino-tensors (Rarita-Schwinger)} . It makes explicit the fact that these functions do not depend on the choice of the heavy quark current and refer only to the light quarks:
\begin{eqnarray}\label{tilde-xi}
\tilde{\xi}(v,v')=~<j'j_z'v'~|~j j_z v>
\end{eqnarray} 

The light cloud state in motion $|~j j_z v >$ is obtained as the result  of a Lorentz boost on rest frame states. It is labelled by the spin at rest with projection along one direction $Oz$. This construction of states implies a covariance property of the overlaps by composition of the Lorentz boosts and the additional Lorentz transformation (see our Lorentz papers for details, for instance eq.(24) of \cite{LorentzI}). Also, from this definition through overlaps, and using simply completeness, the existence of sum rules for currents is obvious.

We call this the "natural" definition of the Isgur-Wise functions, because it does not refer to any particular definition of the quark current and it treats all spins on the same foot. We give them a homogeneous identical notation by a $\xi$ with possible indices of various sorts. This homogeneous notation is justified by the fact that one can give them a common treatment. But to avoid confusion, we signal the difference with the usual definitions by the tilde. These Isgur-Wise functions defined in an homogeneous way for all spins present a possible difference with the standard ones, by a kinematical factor. The relation to the standard definitions of Isgur-Wise functions in particular cases is very easy to establish, and examples are given in the subsection \ref{conversion}. In addition, in particular discussions, we use a specific notation ($\tilde{\tau}$) for $j=0 \to j=1$ transitions for sake of clarity.

\subsection{The BT model as an implementation of Falk's formulation}

The Bakamjian-Thomas model for current matrix elements, as defined in the initial formulae of our original paper  \cite{covariant}, section 1, appears in the heavy quark limit as the particular case of eq. (\ref{falkPost}) where the above overlap is given by the scalar product of ordinary wave functions. Indeed, as stressed in \cite{exact}, the heavy quark velocities coincide with the hadron velocities ; then, the Wigner rotations relative to the heavy quark reduce to the identity, while the heavy quark spinors now depend only on these hadron velocities and they can be trivially factorised. After having thus factorised the heavy quark current in the formula for BT, we remain with a quantity relative only to light quarks as in eqn.(\ref{tilde-xi})\footnote{ \label{factor} Everything in $\tilde{\xi}$ is relative to the light quark. In the paper \cite{exact}, a possible dependence on the heavy quark spin $s_1$ was retained. This is however trivial if there is no dependence of the mass operator on the heavy quark spin : the remaining $s_1$ dependence is now trivially extracted by factorising the above CG coefficients between the heavy quark spin and the light cloud spin $j$. We adopt here this assumption for the mass operator, which is valid for standard spin forces, since those related to the heavy quark are suppressed by heavy mass factors} .

One can now write the overlap of eq.(\ref{tilde-xi}) in terms of the the wave function in motion for light quarks denoted by $\psi$:
\begin{eqnarray} \label{tilde-xiBT}
\tilde{\xi}(v,v')=~<\psi^{j' j_z',v'}_{n'} \mid \psi^{j j_z,v}_{n}>
\end{eqnarray}
where $n,n'$ denote the eigenstates of the mass operator from which the $\psi$'s are deduced. The angular momentum quantum numbers are still kept explicit for the moment to ensure continuity with the initial formulae, although they are redundant with $n,n'$). These  $n,n'$ are internal labels .  The wave functions $\psi$ are obtained by an elementary Lorentz transformation of momenta and spins from the internal wave functions, eigenstates of the mass operator. The relation between the two types of wave functions is described in section \ref{first}. The scalar product is taken by integration over momenta and contraction on spins \footnote{To simplify the notation, we have included the Jacobian factors of the paper \cite{covariant} in the wave function in motion. See eqn. (\ref{psi}).}
The states being obtained from the states at rest by a Lorentz boost, this completes the demonstration that BT in the heavy quark limit is a simple realisation of the above Falk assumptions \cite{falk}, with the light cloud being represented by a fixed number of light quarks. This construction by boosts implies a covariance of the overlaps, according to the general demonstration referred to in the preceding subsection, to be recalled in more detail in section \ref{covariance}). 



\subsection{Extraction of the center-of-mass motion in the BT model}

A further and essential step described in our paper \cite{exact} (section 3),
at least in a limited expansion in velocities, is to extract explicitly the dependence on the center of mass motion of the hadrons and to formulate the initial matrix elements in terms of new matrix elements between internal wave functions. All effects of hadron motion are included in operators inserted inside the scalar product of these internal states. Then, the argument of both wave functions is now the set of the integration variables themselves without Lorentz transformation \footnote{This is in contrast with our usual expressions for Isgur-Wise functions, where, for instance, the argument of the internal wave functions depends explicitly on velocities through $(p.v)^2,(p.v')^2$, $\vec{p}$ being the integration variable}. We then denote the internal states or states at rest by round brackets $(|$ or $|)$ to avoid confusion. From now on, when not stated differently, "states" denote these internal states, and the term "matrix elements" denote the final ones with round brackets. We can then write the Isgur-Wise functions defined in eq. (\ref{tilde-xi})~(with subindices relative to these internal states): 
\begin{eqnarray} \label{tildeiBT0}
\tilde{\xi}_{n~\to~n';~v,~v'}=(n'|O(v,v') |n)
\end{eqnarray}
The internal states have some definite spin state $j,j_z$ (for instance $1/2,+1/2$ for the light quark of a ground state meson), and of course there are relations between $\tilde{\xi}$ relative to different $j_z$, but we do not need to specify them presently. For simplicity, we choose collinear velocities along the axis $Oz$ which is the one for the spin projection. Of course this is a strong restriction on the possibility of exploiting the covariance : it will be restricted to a collinear subgroup. This restriction is removed in the end of the paper to demonstrate the full covariance (section \ref{covariance}). Full covariance allows to relate the Isgur-Wise functions for completely arbitrary velocities $v,v'$ to the ones for collinear $v,v'$, thanks to suitable Lorentz transformations, which imply Wigner rotations. 

The step of "extracting" hadron motion is the one described in section \ref{first} of the paper for infinitesimal velocities. Finite velocities will be treated in section \ref{IW}. The operator describing the motion, $O(v,v')$, comes out to be the exponential of a very simple operator ${\cal O}$ corresponding to infinitesimal velocity. Both will be explicited below (section \ref{IW}). Extension to general HQET is simply obtained on replacing ${\cal O}$ by some generator of Lorentz transformations along $Oz$.

\subsection{Conversion to usual notations for Isgur-Wise functions} \label{conversion}

We have deliberately chosen a general definition of the Isgur-Wise functions, denoted as $\tilde{\xi}$, which then does not coincide, most often, with the usual ones, the $\xi$'s, representing the full hadronic and manifestly covariant current (i.e. with the heavy quark current factor included), and which are defined through certain conventions, in particular a choice of certain Lorentz covariants (for instance $(v^{\mu}+v^{\mu'})\xi$ for the meson current). 

Performing the conversion of the $\tilde{\xi}$'s to these usual $\xi$'s, a basic dissymmetry appears between integer and half integer $j$, or even and odd number of quarks, and finally between all the values of $j$ \footnote{Such a dissymmetry remains in the paper of Falk because the light quark spin in overlaps remains  described in the Rarita-Schwinger formalism.}. 
This is why, wanting to be general, we stick here to the two-dimensional representation of spin states, as in the initial formalism \footnote{Note however: a difference with our initial two-dimensional spin formalism is that we extract from the beginning the heavy quark current, and work only with the light quarks, as in our Lorentz papers. In short, we try now to take advantage both of the two-dimensional description of spin and of the approach of Falk through light quark overlaps.}.



Note that the conversion implies at the same time a possible difference in the normalisation of states, and possible additional factors in particular conventions, like a $\sqrt{2}$ in the definition $\zeta=\sqrt{2}~\tau_{1/2}$ or similarly $\tau=3~\tau_{3/2}$. As to normalisation, our normal choice is $\delta(\vec{P'}-\vec{P})$, which has the advantage of leading directly to the usual non relativistic limits.

The case $j=0$ is immediate and the conclusion is that $\xi=\tilde{\xi}$. Indeed, converting the l.h.s. to the $\delta(\vec{P'}-\vec{P})$ normalisation :
\begin{eqnarray}
\bar{u}_s' (v') u_s (v)~\xi(w)=\bar{u}_{s_{1}'}(v') u_{s_{1}}(v)~\tilde{\xi} (w)
\end{eqnarray}
with the additional fact that, since $j=0$, $s'=s'_{1}=s_{1}=s$ (spin is conserved in the heavy quark limit). The velocities of the heavy quark are the same as those of the heavy hadron because one works in the heavy mass limit.
 
To illustrate the procedure in a less immediate case, we treat another example of conversion in the case of a ground state meson $J^P=0^-$($j=1/2$). In standard covariant terms, the scaling Isgur-Wise function is defined through :
\begin{eqnarray} 
<J^{\mu}_V>~=(v^{\mu}+v'^{\mu}) \xi
\end{eqnarray}
with the usual definition of $\xi$.

On the other hand, in our terms, as considered in eq.(\ref{falkPost}), one has a factor :
\begin{eqnarray}
\bar{u}_{s_{1}'}(v') \Gamma u_{s_{1}}(v)
\end{eqnarray}
times the new $\xi$, i.e. $\tilde{\xi}$ homogeneously defined in eq.(\ref{tilde-xi})( with the relevant $\Gamma=\gamma_{\mu}$).

But one has also to account for the fact that the normalisation of states should be chosen consistently on both sides. For instance, we can adopt again for the states the non covariant normalisation. The relation is then written :
\begin{eqnarray}
\frac {(v^{\mu}+v'^{\mu})~\xi} {2\sqrt{ v_0 v'_0}}= \bar{u}_{s'_{1}}(v')~\gamma^{\mu} u_{s_{1}}(v)~\tilde{\xi}
\end{eqnarray}
with also the non covariant normalisation for spinors.
Therefore, to relate $\xi$ to $\tilde{\xi}$ amounts to calculate the ratio :
\begin{eqnarray}
\frac{ \bar{u}_{s'_{1}}(v')~\gamma^{\mu} u_{s_{1}}(v)} {(\frac {(v^{\mu}+v'^{\mu})} {2 \sqrt{v_0 v'_0}})}
\end{eqnarray}
 A simple calculation with collinear velocities shows that :
\begin{eqnarray}
\frac{ \bar{u}_{s'_{1}}(v')~\gamma^{\mu} u_{s_{1}}(v)} {(\frac {(v^{\mu}+v'^{\mu})} {2\sqrt{v_0 v'_0}})}=\sqrt{\frac{2}{w+1}}
\end{eqnarray}
whence :
\begin{eqnarray}
\xi=\sqrt{\frac{2}{w+1}}~\tilde{\xi}
\end{eqnarray}
From this, one can deduce the relations between the corresponding slopes or curvatures (second order derivatives) \footnote{One must note that in our papers on Lorentz group analysis of Isgur-Wise functions, the definition of the Isgur-Wise functions is still another one. It is similar to what is considered by Falk in that it implies only the light cloud, like our present one. However, it describes it in terms of covariant polarisation tensors. One extracts from the light overlap an invariant bilinear describing the dependence on the light cloud spin in Rarita-Schwinger notation (scalar $1$ for $0^+$(corresponding to a $\Lambda$ baryon), spin bilinear $\bar{u}' u$ for a spin $(1/2)^+$
(ground state meson),...). For the spin $(1/2)^+$, one finds a factor $\sqrt {\frac {2}{w+1}} \delta_{j_z,j_{z'}}$ with respect to  our present "natural" definition; it happens then to coincide with the usual manifestly covariant definition from the hadronic current $v^{\mu}+v^{\mu'}$. Whence again a $1/4$ additional term to the slope with respect to our $\tilde{\rho}^2$}
\begin{eqnarray} \label{diffrho2}
{\rho}^2=\tilde{\rho}^2+\frac{1}{4}
\end{eqnarray}

\begin{eqnarray}\label{diffsigma2}
\sigma^2= \tilde{\sigma^2}+\frac {\tilde{\rho}^2}{2}+\frac {3}{16}
\end{eqnarray}
.   

Let us stress again that the use of the new definition of the Isgur-Wise wave functions $\tilde{\xi}$'s instead of more usual ones, like the $\xi$ for mesons, has a big advantage in a general discussion
which is the main aim of the present paper. Indeed, in this respect, having an homogeneous definition for all spins affords a very important simplification ; a large set of demonstrations can be performed without specification of spin.


\subsection{Plan of the paper}

Our first aim here is to present explicit and simple expressions for the {\it slopes} of the IW functions for the ground state and the transitions, obtained in terms of the internal wave functions and a simple differential operator ${\cal O}$, for any number of light quarks. 

A general demonstration that the BT matrix elements satisfy a full set of forward OPE sum rules for
 $m_Q \to \infty$ (namely inclusive duality sum rules in the spirit of the paper by Isgur and Wise \cite{dualityIsgur}) has been given in the first part of our paper on duality \cite{exact}
for an arbitrary number of light quarks (for non-forward sum rules see our paper \cite{nousUraltsev}). 
Our present calculation displays in a manifest form the same duality property at the level of relations between slopes of Isgur-Wise functions-namely the Bjorken sum rule- as simple relations between powers of an operator, with explicit expressions of these slopes in terms of matrix elements between internal wave functions (see for instance eqn. (\ref{SR}) below). 
This operator formulation is then extended to the full 
Isgur-Wise function $\xi(w)$ itself and to the expression of higher derivatives (section \ref{IW}, subsection \ref{exp} and \ref{higher}).

Up to there, we are working with the BT quark models. Starting from section \ref{HQET}, 
a second aim is to generalise the formulation to HQET, which is rather straightforward,  and to show that, as regards heavy quark symmetry, BT quark models are just a natural realisation of the general case with a particular choice 
of Lorentz group generators in the space of light quanta : namely the generators for free light quarks.
This allows a general treatment of a large set of HQET sum rules, which are immediately valid for the BT models.

Finally, the questions about the covariance of the approach in the case of BT are discussed in detail in section \ref{covariance}.
 
\section{Recalling statements of a previous paper and announcing results} \label{recall}

In fact, in section 3 of our paper on duality in the BT approach \cite{exact}, a general, explicit, expression for any number of light quarks has been given for the transitions to $L=1$ states.  A calculation of $\rho^2$ limited to mesons was also summarised. It resulted in a new demonstration of the Bjorken sum rule for mesons, physically transparent in the sense of emphasizing the separate spatial and Wigner rotation contributions, and mathematically trivial as an identity between operators. We first recall these results in the present section.

We then display them in more detailed, general and rigorous way in sections \ref{first}, 
\ref{generalrho2} and \ref{IW}. 

In the following, for intuitiveness and proximity to the non-relativistic wave mechanics, unless specified, we use the 3-dimensional velocity and $\vec{v}$ to denote it, and we then express everywhere the 4-dimensional $v^0$ as $1/\sqrt{1-(v^z)^2} \simeq 1+\frac{1}{2} (v^z)^2$.

The main observation is that, with hadron velocities in the $z$ direction, the same hermitian operator, acting on each light quark \footnote{We write space-time indices as superindices. Lower indices will denote the labelling of light quarks. The final state is signalled by a prime $'$.}:
\begin{eqnarray} \label{calO} 
{\cal O}=-\frac{z p^0+p^0 z}{2}+\frac{1}{2} \frac {(\vec{\sigma} \times \vec{p})^z} {p^0+m}
\end{eqnarray}
gives at the same time (eqs. (28)(34)(35) of \cite{exact} \footnote{The numbering refers to the published version, which differs somewhat of the {\it hep} one.}) :

1) the slope  with respect to $v_z'-v_z$ of the transition from the ground state to the excited states through the corresponding matrix element of ${\cal O}$.

2) the slope of the ground state IW function with respect to $(v_z'-v_z)^2$  through the diagonal matrix element of its square ${\cal O}^\dagger{\cal O}={\cal O}^2$. Conventionally, one defines this slope  with respect to $-(w-1)\simeq-\frac{1} {2}(v_z'-v_z)^2$. Then it is seen that, for mesons,
the standard $\rho^2$ is the matrix element of ${\cal O}^2$ + $\frac{1}{4}$ ; this  $\frac{1}{4}$ corresponds to the term in eq. (\ref{diffrho2}). If on the other hand one uses our uniform definition of Isgur-Wise functions, one has simply $\tilde{\rho^2}={\cal O}^2$, a result which will be  shown to hold for baryons as well.

${\cal O}$ appears as the sum of two terms, corresponding respectively to space and spin (Wigner) transformations of the wave functions, which we denote hereafter as ${\cal O}^{z,T}$, referring to their respective orbital angular momentum structure (in the first term it is governed by $z$, in the second by $\vec{p}~^T$). When there are several light quarks, one has to take the sum over one similar operator for each light quark. We then note ${\cal O}_k$ the operator acting on the quark $k$, and ${\cal O}$ will denote the sum, 
\begin{eqnarray} 
{\cal O}=\sum_k {\cal O}_k
\end{eqnarray}
${\cal O}$ is obviously hermitian as ${\cal O}_k$.

We denote generically the matrix elements for a transition from the ground state to an $L=1$ excited state as:
\begin{eqnarray} 
<n,L=1,v'|0,v>=(v'-v)\tilde{\tau}_n +...(\mathrm{higher~orders~in~3-velocities}).
\end{eqnarray}
Therefore the  $\tilde{\tau}_n$'s are the above slopes 1) (the notation $\tilde{\tau}_n$ is meant to recall the $\tau_{1/2,3/2}$'s of $L=1$ heavy mesons ). 

Then the first statement is :
\begin{eqnarray} \label{tau}
\tilde{\tau}_n=(n|{\cal O}|0)
\end{eqnarray}
while the second statement is :
\begin{eqnarray} \label{rho2}
\tilde{\rho}~^2=(0|{\cal O}^2|0)
\end{eqnarray}
The latter was demonstrated only for one light quark. The general demonstration for any number of quarks is given in the present paper, see sections \ref{generalrho2} and \ref{higher}.


The sum of the squares of slopes 1) equates the slope 2) by the following chain of equalities to be recalled below :
\begin{eqnarray} \label{SR}
\sum_n |\tilde{\tau_n}|^2=\sum_n |(n|{\cal O}|0)|^2=(0|{\cal O}^2|0)=\tilde{\rho}~^2
\end{eqnarray}
where in the first step one uses completeness of internal eigenstates.

This is the restricted Bjorken sum rule for slopes in its simplest form, with a natural definiton of slopes, denoted by a tilde, corresponding to the general definition of eq. (\ref{tilde-xi}). 
The closure runs of course in principle on all states in the Hilbert space of internal states, but the sum can be restricted to the states with $L=1$, since ${\cal O}$ is itself a $L=1$ operator, and other matrix elements vanish.
Then another remarkable fact is that in the sum rule the Wigner rotation contribution does not interfere with the spatial contribution : they add in squares. This is explained below in general terms (section \ref{general}).

\section{Expansion at first order in external velocities $v^z,v^{'z}$: transitions to $L=1$ states} \label{first}

Let us first show how one establishes the above equation (\ref{tau}). After extraction of the heavy quark current, the remaining matrix element appears as the scalar product of wave functions in motion, which are themselves the result of a Lorentz boost acting on the internal wave functions-these ones being eigenstates of a standard spectroscopic Hamiltonian \footnote{We label momentaneously the quarks by $i$, with $i>1$ for light quarks, instead of our usual $k$ to avoid confusions.},
\begin{eqnarray} \label{psi}
\psi^{v}(\vec{p_i})=\left(\prod_{i>1}\sqrt{\frac {p_i.v}{ p_i^0}} D_{i}^{1/2}\right) \phi ^{}(\vec{k_i}(\vec{p_i})),
\end{eqnarray}
where $\vec{k_i}(\vec{p_i})$ is the Lorentz transformation of $\vec{p_i}$ leading to the hadron rest frame(see eqn. \ref{KP} below). The Jacobian factors  $\prod_{i>1}\sqrt{\frac {p_i.v}{ p_i^0}}$ corresponding to the change of variables between the $\vec{p_i}$ and the $\vec{k_i}(\vec{p_i})$ are included in the moving hadron wave function, so that the matrix element is simply the overlap as it should:
\begin{eqnarray}
   <\psi^{v}_{n'}(\vec{p_i}) \mid \psi^{v}_{n}(\vec{p_i})>
\end{eqnarray}
with integration on momenta $\vec{p_i}$ and contraction on spin indices, and $n,n'$ being the additional label of the states. The $D_{i}^{1/2}$'s act on the spin part of $\phi ^{j\mu}$.  In the practical case, with $v$ (or $v'$) along the $Oz$ frame, one has :
\begin{eqnarray} \label{KP}
k^z_i (\vec{p_i})=\frac{p^z_i-\beta p^0_i}{\sqrt{1-\beta^2}},~\vec{k^T_i}=\vec{p^T_i}
\end{eqnarray}
$\beta$ being the 3-velocity $\beta=v^z$ \footnote{Recall that we have decided for intuitiveness to work with the 3-velocity and denote it as $\vec{v}$.} and $p^0_i=\sqrt{\vec{p_i}^2+m^2_i}$.
The task is now to look for the terms of first order in the velocities $v^z,v^{'z}$. 
For infinitesimal velocity $v^z$, one has a sum over all the light quarks of all the effects relative to one quark, each one being of order one (the matrix element is $0$ or $1$ at $v^z,v^{'z}=0$ because of orthogonality of internal states) .
 
$D_{i}^{1/2}$ can be expanded into :
\begin{eqnarray} \label{D}
D_{i}^{1/2}=1-i~v^z \frac{1}{2} \frac {(\vec{\sigma_i} \times \vec{p^T_i})^z} {p^0_i+m_i}+...
\end{eqnarray}
and :
\begin{eqnarray} \label{kp}
k^z_i=p^z_i- v^z p^0_i+...,~\vec{k^T_i}=\vec{p^T_i}
\end{eqnarray}
From eq. (\ref{kp}), one can derive a differential expression for the wave function $\phi ^{j\mu}(\vec{k_i}(\vec{p_i}))$ :
\begin{eqnarray} 
\phi ^{}(\vec{k_i}(\vec{p_i}))=\phi ^{}(\vec{p_i})-v^z \sum_i p^0_i  \frac{\partial \phi}{\partial p^z_i} +...
\end{eqnarray}

Finally, the factor $\sqrt{\frac {p_i.v}{ p_i^0}}$ can be expanded as:
\begin{eqnarray} \label{jacob}
\sqrt{\frac {p_i.v}{ p_i^0}}=1-v^z\frac{p^z_i}{2~p^0_i}+...
\end{eqnarray}

Whence the total effect of the infinitesimal Lorentz transformation 
\begin{eqnarray}
\psi^{v}(\vec{p_i})=\phi(\vec{p_i})-v^z \sum_{i=2,...} \left(p^0  \frac{\partial \phi}{\partial p^z}+\frac{p^z_i}{2~p^0_i} \phi+i~\frac{1}{2} \frac {(\vec{\sigma_i} \times \vec{p^T_i})^z} {p^0_i+m_i}\right)\phi +...
\end{eqnarray}
The same equation can be applied to the final state wave function with the substitution $v^z \rightarrow v^{'z} $, but, as to the spatial part, this does not lead immediately to what is written in  eq. (\ref{tau}) : namely for the transformation of the product $\phi^{'*} \phi $ one gets  the sum of the term coming from $\phi$ and the one from $\phi^{'*}$, but this does not exhibit the expected factor $v^{'z}-v^z$ for the spatial part of the transformation, while the Wigner rotations contribution is automatically $\propto (v^{'z}-v^z)$:
\begin{eqnarray}
\psi^{v',*}_{}(\vec{p_i}) \psi^{v}_{}(\vec{p_i})&=&\phi^{'*} \phi\nonumber \\&-& \sum_{i=2,...}(v^z(p^0_i \phi^{'*} \frac{\partial \phi}{\partial p^z_i}+\frac{p^z_i}{2~p^0_i}\phi^{'*} \phi)+v^{'z}(p^0_i \frac{\partial  \phi^{'*} }{\partial p^z_i }\phi+\frac{p^z_i}{2~p^0_i}\phi^{'*} \phi))+\nonumber \\
&&\mathrm{contribution~of~the~Wigner~rotations}+...
\end{eqnarray}
It is only after considering the scalar product and performing an integration by parts that one exhibits the expected factor. One ends with \footnote{Here we reestablish the indices $n,n'$} :
\begin{eqnarray} \label{v'msv}
   <\psi^{v'}_{n'}(\vec{p_i}) \mid \psi^{v}_{n}(\vec{p_i})>=\delta_{{\cal I},{\cal F}}+ (v^{'z}-v^z) \sum_{i=2,...}( \frac{1}{2} (-p^0_i \frac{\partial  \phi^{'*} }{\partial p^z_i }\phi+p^0_i \phi^{'*} \frac{\partial \phi}{\partial p^z_i}))\nonumber \\+i~(v^{'z}-v^z)  \sum_{i=2,...} \frac{1}{2} \frac {(\vec{\sigma_i} \times \vec{p^T_i})^z} {p^0_i+m_i}+...
\end{eqnarray}

Here, $\delta_{{\cal I},{\cal F}}=1$ for identical internal initial and final states (${\cal I}={\cal F}$), and $0$ otherwise. For the inelastic case, the transition is obviously from $L=0$ to $L=1$ states.

Rewriting $i\frac{\partial}{\partial p^z_i}=z_i$, one recognizes the operator ${\cal O}$ of eqn.(\ref{calO}): in fact, there appears the sum on $i$ of identical operators each given by eqn.(\ref{calO}) acting on one light quark $i$. As said above, ${\cal O}$ then denotes the sum :
\begin{eqnarray} \label{calOi}
{\cal O}=\sum_i {\cal O}_i
\end{eqnarray}

One also recognizes in the eqn. (\ref{v'msv}) the eqn. (\ref{tau}), except for an additional phase $i=\sqrt{-1}$ which is irrelevant.  


\subsection{Decomposition of ${\cal O}$ and structure of $L=1$ final states near $w=1$ for $j=0$}\label{decomposition}

${\cal O}$ is obviously the third component of a vector under the full $SO(3)$ group ($SO(3_{space})\times SO(3)_{spin}$), therefore it bears a total angular momentum $j_{\cal O}=1$. Then, the final states produced near $w=1$ have automatically $j'=1$ if $j=0$, which is an expectation from HQET (see reference \cite{youssefmir}), although the HQET statement is stronger as it applies to any $w$ (see below, subsubsection \ref{selection}, for a proof for arbitrary $w$ valid for both HQET and the BT approach). 

Moreover, it appears as the sum of two contributions for each light quark: a part "longitudinal" in space,
 \begin{eqnarray} \label{Oz}
{\cal O}^z=-\frac{z p^0+p^0 z}{2}~(m_z=0),
\end{eqnarray}
coming from the Lorentz transformation of space or momentum in the wave function and one which is "transverse" :
\begin{eqnarray} \label{0T}
{\cal O}^T=\frac{1}{2} \frac {(\vec{\sigma} \times \vec{p})^z} {p^0+m}~(m_z=\pm 1) 
\end{eqnarray}
coming from the transformation of spin (Wigner rotation)\footnote{\label{indices} One must distinguish the indices for orbital motion (as opposed to spin), from the indices corresponding to tensor behavior under the full $SO(3)$ space. To display this distinction, we label the latter as $1,2,3$, as opposed to $x,y,z$.}. For several light quarks, one has for each one \footnote{We now return to the labelling of light quarks by $k$.}:
\begin{eqnarray} \label{OzT}
{\cal O}_k={\cal O}^z_{k}+{\cal O}^T_{k}
\end{eqnarray}
All the contributions, ${\cal O}^{z,T}_{k_2}$ and ${\cal O}^{z,T}_{k_3}$ commute for $k_2 \neq k_3$ ; on the other hand, ${\cal O}^{z}$ and ${\cal O}^{T}$for the same light quark do not commute with each other because of the denominators in $p^0+m$ in ${\cal O}^{T}$ which contain $p^z$, while ${\cal O}_{z}$ contains $z$. Moreover, ${\cal O}^z$ is symmetric in space coordinates $2,3$, with no spin dependence, while ${\cal O}^T$ is symmetric in combined space and spin exchange for each couple of light quarks, $s_2,p_2 \leftrightarrow s_3,p_3$.


This analysis displays the structure expected for the final states, assuming a $j=0$ initial ground state with $s=0,l=0$ : \footnote{As is well known, in baryons, the spin forces may lead to more complicated structures. However, in the static limit, $j$  and $s$ are conserved, because the spin interactions involving the heavy quark vanish, and those relative to the light quarks commute with $|\vec{S}_{23}|^2=s(s+1)$. See also footnote \ref{factor}.} . The final states should be $l=1$ with either $s=0$ or $s=1$ generated respectively by ${\cal O}^z$ or ${\cal O}^T$ (the latter cannot generate $s=0$ since $(s=0|\vec{\sigma}_i|s=0)=0$). As to symmetry, due to the symmetry of the ground state with respect to space labels $2 \leftrightarrow 3$, and the symmetry in space of ${\cal O}^z$, this operator generates the $"\lambda"$  type baryons, therefore finally it generates the $"\lambda,s=0"$ type. On the other hand, since ${\cal O}^T$ generates the symmetric spin state $s=1$ from the antisymmetric $s=0$, and given the symmetry of the operator under $s_2,p_2 \leftrightarrow s_3,p_3$, it generates a state antisymmetric in space, therefore finally the "$\rho,s=1$" type.

\subsection{Non-relativistic expansion in internal velocities. Retrieving the dipole formula. Hierarchy of transitions from a $j=0,l=0,s=0$ initial state}

In the non-relativistic (NR) expansion in internal velocity, i.e. $|\vec{p_k}| << m_k$ ($\frac {v}{c}<< 1 $), one gets 
\begin{eqnarray} \label{ONR}
{\cal O} \simeq \sum_k m_k z_k.
\end{eqnarray}
This is just the dipole operator, except for the mass factor. 

As to higher orders in this expansion of ${\cal O}$, 

1) the factors $p^0_k$ in front of $z_k$ in eqn.(\ref{Oz}) induce corrections of relative order $O(\frac {v^2}{c^2})$ with respect to $m_i$, but they do not induce any change of the $O(3)$ structure of the transition operator

2) the Wigner rotation term eqn. (\ref{0T}) is $O(\frac {v}{c})$, while the spin independent term, ${\cal O}^z$, the main term, is of order  $O(\frac {c}{v})$, since $m z$ is $O(\frac {c}{v})$. Therefore, its contribution is of relative order $O(\frac {v^2}{c^2})$.
If a transition is induced purely by the Wigner rotation, its rate is then $O(\frac {v^4}{c^4})$ with respect to the main one. When this expansion has some validity, it is then expected that the transitions generated by the Wigner term are much smaller than those generated by the spatial term. 

Of course, this is relevant if the transitions induced by these two respective terms lead actually to different final eigenstates. This is the case for transitions from  baryons in the ground state $j=0$, $s=0$. The two terms ${\cal O}^{z,T}$ being respectively $s_{\cal O}=0,1$ are inducing transitions to $L=1$ states with respectively $s=0$ and $s=1$, which are energy eigenstates. This is the case for heavy $\Lambda$ with space function symmetrical or antisymmetrical in the two light quarks ($"\lambda",s=0$ or
$"\rho",s=1$ types). Whence an expected suppression of the "$\rho$" type with respect to "$\lambda$"type.  This type of spatial symmetry is automatically induced by the symmetry structure of ${\cal O}$ (see above). A quantitative calculation of transitions to $s=1$ is given in the subsection \ref{analysis}.

Differently, in the heavy meson case, starting from the $j=s=\frac {1}{2}$ ground state, one ends with final states which are eigenstates of $j$, with $j=\frac {1}{2},\frac {3}{2}$, but  which are not eigenstates of spin $s$. Therefore, the two terms ${\cal O}^{z,T}$ are contributing to both types of final states. 



\section{General calculation of $\tilde{\rho}^2$ and Bjorken sum rule for an arbitrary number of light spectator quarks. A specific sum rule for transitions to $s=1$ final states} \label{generalrho2}

With expansion in powers of the hadron velocities, one is led to matrix elements of powers of ${\cal O}$, and for average over the ground state, the first non trivial one is $\tilde{\rho^2}$ (order $(v_z'-v_z)^2$) which equates the average of ${\cal O}^2$  (eqn. (\ref{SR})). 
As stated in the old paper \underline{Exact duality...} \cite{exact}, the calculation of $\tilde{\rho}^2$ appeared more involved than the first order, and the result was given only for one light quark.

However, a simple remark allows to get immediately the generalisation to an arbitrary number of light quarks. In fact, one notices that, since $(v_z'-v_z)^2$=$v_z^2+{v'}_z^2
-2v_z{v'}_z$, to extract the coefficient of $(v_z'-v_z)^2$, one can disregard the terms of second order $v_z^2$, ${v'}_z^2$ and look only for $v_z{v'}_z$ terms. But the latter terms are generated by the infinitesimal Lorentz transformations at first order in $v_z$ or ${v'}_z$, acting on the initial and final state respectively, both given by the action of only one power of ${\cal O}_k$ on each respective light quark. They are then given by the sum $(\sum_k \sum_{k'} {\cal O}_k {{\cal O}_{k'}}^*)={\cal O}^2$  and we are led to : 
\begin{eqnarray} \label{rho2prime}
\tilde{\rho}^2=(0|~\sum_{k',k}{\cal O}_{k'} {\cal O}_k~|0)=(0|~(\sum_k{\cal O}_k)^2|0)=
(0|~{\cal O}^2|0)
\end{eqnarray}  
The Bjorken sum rule\footnote{here written for slopes in "natural" notations, eqn.(\ref{SR}); it could take various forms when translated into covariant notations.} now follows for the case with any number of light quarks, by inserting the complete set of states between the two ${\cal O}$ 's , without any assumptiom on a separability of the internal Hamiltonian into separate contributions for each light quark:
\begin{eqnarray} \label{rho2Bj}
\tilde{\rho}^2=(0|~{\cal O}^2|0)= \sum_n (n|{\cal O}|0)^2=\sum_n |\tilde{\tau_n}|^2
\end{eqnarray}  
just as in section \ref{recall}. As said above, both members of eq. \ref{SR} decompose into a sum of contributions, one from the space part of ${\cal O}$ and one from Wigner rotations. Indeed, in average of the square of ${\cal O}$, due to the fact that the space part $-i \frac{z p_0+p_0 z}{2}={\cal O}^z$ has a structure $m_z=0$ while the Wigner part ${\cal O}^T$ has $m_z=\pm 1$, any crossed term vanishes on integration with $\varphi_0$. 
\vskip 0.5cm
Therefore, one ends with the decomposition:
\begin{eqnarray} \label{rho2sep}
\tilde{\rho}^2&=&(0|~|{\cal O}^z|^2+|{\cal O}^T|^2|0)\nonumber \\&=&(0|~|\sum_k  (-\frac{z_k p_k^0+p_k^0 z_k}{2})|^2|0)+(0|~|\sum_k\frac{1}{2} \frac{(\vec{\sigma} \times \vec{p}_k)_z}{p_k^0+m_k}|^2|0) 
\end{eqnarray}
and the same for $\sum_n |\tilde{\tau_n}|^2$ :
\begin{eqnarray} \label{tau2sep}
\sum_n |\tilde{\tau_n}|^2=\sum_n|(n|~\sum_k  (\frac{z_k p_k^0+p_k^0 z_k}{2})~|0)|^2+\sum_n|(n|\sum_k\frac{1}{2} \frac{(\vec{\sigma} \times \vec{p}_k)_z}{p_k^0+m_k}|0)|^2  
\end{eqnarray}
Note that in eqn.(\ref {tau2sep}), the vanishing of mixed longitudinal-transverse terms does not always hold for each term separately, but only in the sum on $n$. For instance, one has to sum over $j=1/2$ and $j=3/2$ states for $L=1$ mesons (see eq. (32) of \cite{exact}).


\subsection{Non relativistic limit of $\tilde{\rho^2}$ and intuitive interpretation of the spatial part} \label{sectionNR}
In the NR limit $\vec{p_i} << m_i$, the Wigner rotation contribution tends to be negligible,-of order $v^2/c2$, as already explained, while the main term is $c^2/v^2$, therefore :
\begin{eqnarray}
\tilde{\rho^2} \simeq <0\mid(\sum_k m_k z_k)^2 \mid 0>=<0\mid(\sum_{k,k'} m_k m_{k'} z_k z_{k'}) \mid 0>
\end{eqnarray}
A remarkable fact is that in this approximation, $\tilde{\rho^2}$ depends only on the coordinate of the center of gravity of the light quarks and not on relative coordinates like $\vec{r_2}-\vec{r_3}$ for baryons. On the other hand, there are in general "crossed" contributions from averages of e.g. $z_2 z_3$ because nothing forbid to have a $l_2=l_3=1$  component in the $j=0,~l=0$ wave function.

Away from the non relativistic limit, Wigner rotations seem practically to have still a minor effect. The main effect is the substitution of the light quark masses $m_i$ by the quark energies in the spatial part. This is a very simple and intuitive picture.
  

\subsection{Analysis of $\rho^2$ in terms of $L=1$ intermediate states for $\Lambda_b$ baryons; a sum rule for transitions to $s=1$ } \label{analysis}

We consider the case of $j=s=0,J^P=\frac{1}{2}^+$ heavy baryons, which is simpler than that of mesons \footnote{This may seem paradoxical, but one must be aware that here, the number of quarks is not relevant; what matters is the various angular momenta of the light quark cloud.}. Recall that in this case $\tilde{\rho}^2$ equals the usual $\rho^2$ : there is no conversion factor.
We could consider as the still simplest case the heavy quark-light diquark model of $J^P=\frac{1}{2}^+$ baryons, but we aim to illustrate the advantage of the method for several light quarks, therefore we choose the two light quark case.

We note that the two perfect squares in the expression of $\tilde{\rho}^2$ eqn.(\ref{rho2sep}), correspond, in the Bjorken sum rule, to the two different categories of $L=1$ heavy baryon intermediate eigenstates, described above, which have respectively $s=0$ or $s=1$, correlated with a symmetric or respectively an antisymmetric spatial wave function for the case of isospin $I_{2,3}=0$($\Lambda_b$ baryons).
Indeed, having separated the two contributions as squares $(0|({\cal O}^{z,T})^2|0)$, we can insert the two types of states inside the squares:
\begin{eqnarray}
(0|({\cal O}^{z,T})^2|0)=\sum_n (0|{\cal O}^{z,T}|n)(n|{\cal O}^{z,T}|0).
\end{eqnarray}
One sees immediately that intermediate states $n$ with $s=0$ contribute only to $(0|({\cal O}^z)^2|0)$ and $s=1$ states contribute only to $(0|({\cal O}^T)^2|0)$ since only ${\cal O}^{T}$ contains a spin operator.

It is not difficult to perform an explicit calculation of this $s=1$ intermediate states contribution. We start from a little more explicit form of eqn. (\ref{tau2sep}), with explicitation of the spin wave function of the $|0)$ state, $\chi^0$:

\begin{eqnarray} \label{wigner2dim}
\rho^2_{Wigner}=\int \prod_{k=2,3} d^3 \vec{p}_k
\frac{1}{4} (\chi^0)^{\dagger}|\frac {(\vec{\sigma}_2 \times \vec{p}_2)_z}{p^0_2+m}+\frac {(\vec{\sigma}_3 \times \vec{p}_3)_z}{p^0_3+m}|^2\chi^0 |\varphi_0|^2
\end{eqnarray}

We can first reexpress the matrix element $(\chi^0)^{\dagger}...\chi^0$ as a square:
\begin{equation} \label{spin}
|(\frac {(\vec{\sigma}_2 \times \vec{p}_2)_z}{p^0_2+m}+\frac {(\vec{\sigma}_3 \times \vec{p}_3)_z}{p^0_3+m})\chi^0|^2
\end{equation}
and then make explicit the spin wave function $\chi^0$ of the baryon. The spin wave function inside the modulus squared (taking into account the action of the spin operators $\vec{\sigma}_i$) is found to be a linear combination of the two orthogonal $\chi^1_{\pm 1}$. This makes very explicit that we end with $s=1$ intermediate states-or final states in terms of the transition matrix elements $\tilde{\tau_n}$. The orthogonality allows again to write the expression in (\ref{spin}) as a sum of two (identical) squares. We then obtain :
\begin{equation}
|(\frac {p^y_2}{p^0_2+m}-\frac {p^y_3}{p^0_3+m})-i(\frac {p^x_2}{p^0_2+m}-\frac {p^x_3}{p^0_3+m})|^2
\end{equation}
and finally the sum rule for transitions to $n,j=1,s=1 (\rho~\mathrm{type})$ states :
\begin{eqnarray} \label{wignerCov}
\Sigma_{s=1}|(n,j=1,s=1 (\rho~\mathrm{type})|{\cal O}|j=0,s=0)|^2=\nonumber\\\int \prod_{k=2,3} d^3 \vec{p}_k
\frac{1}{4} |\frac {\vec{p}^T_2}{p^0_2+m}-\frac {\vec{p}^T_3}{p^0_3+m}|^2 |\varphi_0|^2=\nonumber \\\int \prod_{k=2,3} d^3 \vec{p}_k
\frac{1}{4} |(\frac {\vec{p}^T_2}{p^0_2+m}-\frac {\vec{p}^T_3}{p^0_3+m})\varphi_0|^2
\end{eqnarray}

Note that this sum rule makes use of the specific structure of ${\cal O}$ in Bakamjian-Thomas models, and therefore it is not a general exact HQET rule unlike the other ones which we are considering in this article (Bjorken sum rule, curvature sum rule, naturality selection rule...). Nevertheless, it may be qualitatively indicative in the real situation. 

It seems in addition that light quarks are less relativistic in baryons than in mesons. Therefore, as explained above, the expectation would be that these $s=1$ states are less copiously produced than the states with same $J^P$ but with $s=0$, since the Wigner operator squared is only a high order relativistic correction to the main spatial one ($O(\frac {v^4}{c^4})$). 

In fact, as seen from the eqns. (\ref{rho2sep},\ref{tau2sep}), the r.h.s. of eqn. (\ref{wignerCov}) gives at the same times an estimate of the rate to $s=1$, $\rho$ type states for $w \simeq 1$ ( up to a factor $\propto w-1$), provided one neglects the difference of masses.
Then, this should be reflected in the respective magnitude of the two contributions to $\rho^2$.

And indeed, a direct calculation of $\rho^2$ with a simple gaussian wave function suggested by Pervin, Roberts and Capstick, \cite{pervin}, shows that the Wigner term contribution to $\rho^2$ is rather small.


\section{${\cal O}$ as the generator of a Lorentz transformation on internal wave functions along $Oz$. The finite Isgur-Wise functions and the higher derivatives.} \label{IW}

\subsection{The Lorentz group structure which underlies ${\cal O}$} \label{general}

(1) As we show now, the above simplification obtained for the second order slope $\rho^2$, leading directly to the expression \ref{rho2}, reflects the group property of the longitudinal Lorentz subgroup. This derives from the fact that the hermitic operator ${\cal O}$ is in fact the infinitesimal generator which describes the effect of the infinitesimal Lorentz transformation on the internal wave functions : indeed, one can rewrite 
:
\begin{eqnarray} \label{lorentz}
\phi 
 =\phi-~i v {\cal O} \phi +...\nonumber \\
\phi'^* 
 =\phi'^*+i~v' {\cal O} \phi'^*+...
\end{eqnarray}
. It must be understood that i) these equations are valid only for matrix elements, so that  integrations by parts are allowed and are required to obtain the present expression of the transformation. This should be more safely expressed in terms of the overlap integrals as done in the previous section; ii) the transformation includes the effect of the Jacobian, and this allows it to be unitary, or its generator ${\cal O}$ to be hermitian as it is evident from equation (\ref{calO}). As to the $D_{i}^{1/2}$'s acting on the spin part of $\phi ^{j\mu}$, they are unitary by themselves. 

(2) It is expected that we deal with a unitary representation of the full Lorentz group, acting in the space of wave functions. It is easily recognised that ${\cal O}$ is the well-known generator of the representation of Lorentz transformations along $Oz$ for an arbitrary number of free light fermions $k=2,3,..$. It is obviously the reason why it describes the motion along $Oz$.

Therefore, it can be embedded in a full Lorentz algebra by considering the other generators for a set of free particles : the generators of Lorentz transformations along $x,y$ (by making $z \leftrightarrow x$ or $z \leftrightarrow y$), which we denote by superindices $1,2$ (with ${\cal O}$ corresponding to $3$) and the three generators of rotations $J^{1,2,3}$. 
These additional generators will be used later and their commutation relations are actually used in the subsection \ref{curvatureHQET}. And the fact that one has this full Lorentz algebra ensures covariance, as discussed in detail in section \ref{covariance}.

\subsection{Exponentiation of ${\cal O}$ and the Isgur-Wise function}\label{exp}

One can exploit the fact that the collinear Lorentz transformations form a commutative group to write an exponential expression of the finite transformation with the infinitesimal generator ${\cal O}$ :
\begin{eqnarray} \label{U}
U(v)=e^{-i \theta(v) {\cal O}}
\end{eqnarray}
where $\theta(v)$ is some function with the property that $\theta(v) \simeq v$  for small $v$, and $U(\theta(v_1)) U(\theta(v_2))=U(\theta(v_1)+\theta(v_2))$.
$U(v)$ is obviously unitary. From the law of composition of velocities 
\begin{eqnarray}  \label{UU}
U(v_1) U(v_2)=U(\frac{v_1+v_2}{1+v_1 v_2}),
\end{eqnarray}
one must have $\theta(v_1)+\theta(v_2)=\theta(\frac{v_1+v_2}{1+v_1 v_2})$.
This functional equation is satisfied by  $\theta(v)=\mathrm{argtanh}(v)$ 
\begin{eqnarray}  \label{Uv}
U(v)=e^{-i~\mathrm{argtanh}(v) {\cal O}}
\end{eqnarray}
Therefore the matrix element representing the IW function can be written as :
\begin{eqnarray} \label{xinn'1}
\tilde{\xi}_{n \to n'}(v,v')=(n'|U^{\dagger}(v') U(v)|n)=(n'| e^{i~\mathrm{argtanh}(v') {\cal O}} e^{-i~ \mathrm{argtanh}(v) {\cal O}}|n)
\end{eqnarray}

A simple calculation shows that  
\begin{eqnarray}
\mathrm{argtanh}(v')-\mathrm{argtanh}(v)= \pm \mathrm{argsinh}(\sqrt{w^2-1})
\end{eqnarray}
with $\pm$ according to the sign of $v'-v$. Let us set $(v'-v)>0$ for definiteness. One ends with :
\begin{eqnarray} \label{xinn'}
\tilde{\xi}_{n \to n'}(v,v')=(n'| e^{i~ \mathrm{argsinh}(\sqrt{w^2-1}){\cal O}} |n)
\end{eqnarray}

Another expression, in terms of the relative velocity ,gives some more intuition and avoids this unpleasant sign convention. Using:
\begin{eqnarray}
\mathrm{argtanh}(v')-\mathrm{argtanh}(v)= \mathrm{argtanh}(\frac{v'-v}{1-v'v})
\end{eqnarray}
one gets :
\begin{eqnarray} \label{xinn'3vel}
\tilde{\xi}_{n \to n'}(v,v')(=(n'| e^{i~ \mathrm{argtanh}(\frac{v'-v}{1-v'v}){\cal O}} |n))=(n'| U(\frac{v'-v}{1-v'v}) |n)
\end{eqnarray}
$\frac{v'-v}{1-v'v}$ has a clear meaning : it is the relative velocity of the final state, as measured in the initial state rest frame. One recognises the law of compositon of velocities. Eqn.(\ref{xinn'3vel}) is a relativistic extension of the well known NR formula:
\begin{eqnarray} \label{xiNR}
\tilde{\xi}^{NR}_{n \to n'}(v,v')(=(n'| e^{i~ k~z} |n))=(n'| e^{i~ (v'-v)~m~z} |n)
\end{eqnarray}
since in the NR limit $\frac{v'-v}{1-v'v} \simeq v'-v$, and ${\cal O} \simeq mz$  for NR internal velocities and for one light quark.  

\subsection{Higher derivatives. Curvature} \label{higher}
This exponential representation allows to obtain simple expressions of the higher derivatives near $w=1$ with the help of ${\cal O}$. One can write a series expansion of eq. (\ref{xinn'}). If $n,n'$ have the same parity, there only survive the even powers of the odd operator ${\cal O}$. Then, for instance for the ground state $n=n'=0$:
\begin{eqnarray} 
\tilde{\xi}(w)=(0|\sum_s (-1)^s \frac {(\mathrm{argsinh}(\sqrt{w^2-1}))^{2s}}{(2s)!} {\cal O}^{2s}|0)
\end{eqnarray}
By further expansion of $\mathrm{argsinh}(\sqrt{w^2-1})$ in series of $w-1$, this gives immediately simple expressions for the derivatives of $\tilde{\xi}(w)$. 

One gets again, in a still more straightforward way, the above relation, eq. (\ref{rho2}) :
\begin{eqnarray} 
\tilde{\rho}^2=(0|{\cal O}^2|0)
\end{eqnarray}
leading to the Bjorken sum rule just by insertion of intermediate $L=1$ states.

Higher order derivatives of order $2s$ ($s>1$) with respect to $w$ are linear combinations of $(0|{\cal O}^{2s}|0)$ and lower powers $(0|{\cal O}^{2m}|0)$ ($m<s$). For instance one finds:
\begin{eqnarray} \label{curvatureO}
\tilde{\sigma}^2=\frac{1}{3}\left (  (0|{\cal O}^4|0)+(0|{\cal O}^2|0)  \right )
\end{eqnarray}.


Relations such as eq.(\ref{curvatureO}) are a possible basis for sum rules beyond the Bjorken one. These sum rules have been demonstated in previous work in HQET and, partly, in the BT quark model approach,-the latter, for mesons. We derive them in subsection \ref{curvatureHQET} in a unified treatment which allows to demonstrate them in BT also for $j=0$ baryons with any number of light quarks.
But before proceeding to this derivation, we want first 1) to recall for sake of clarity the various demonstrations which have been performed, in their respective relation to BT or HQET, and 2) to present a simplified case to illustrate the method .

\subsection{Summary of BT and HQET results for curvature sum rules}

In fact, one has the following results :

1) We have demonstrated a 
sum rules implying the curvature and the transitions to radial excitations in exact HQET, for mesons(\cite{sigmaM}) and also for $j=0$ baryons \cite{sigmaB}. The baryon sum rule is written in next section, subsection \ref{curvatureHQET}, eq. \ref{curvature} and is demonstrated again by a method resting on the present formulation \footnote{In the next subsection, \ref{higherNR}, we use the same method in a NR approximation for illustration, but this assumption is removed in \ref{curvatureHQET}}. 

2) These sum rules are also valid for exact, relativistic BT : this has been demonstrated already for mesons \cite{sigmaMBT}, and it has been checked numerically with a high precision in \cite{unexpected}. Analogously, the HQET baryon sum rule for $j=0$ baryons holds in BT : this latter result is demonstrated below as a direct consequence of the demonstration for general HQET, subsections \ref{curvatureHQET}, \ref{automatic}. 

3) Indeed, it is important to underline that, thanks to the present formulation, the demonstrations in HQET automatically apply to BT. This is the reason why we refer the reader for both to the next section \ref{HQET}. It also shows that the sum rules are valid in BT for any mumber of light quarks.

\subsection{NR limit and NR sum rules} \label{higherNR}

But first, to give a simpler idea of the method to derive sum rules from expressions such as eq.(\ref{curvatureO}), one can take the NR limit as we have done in certain previous sections.
Then, one uses the NR expression of ${\cal O}$, ${\cal O} \simeq \sum_k m_k z_k$ (eqn.(\ref{ONR})) \footnote{Let us underline that here, the NR limit is taken only on internal light quark velocities.  We have not to face the problem of the frame dependence encountered in a fully non relativistic calculation and discussed in ref. \cite{NRbounds}}. Hereafter, in this subsection, we will maintain the notation ${\cal O}$ to denote the non-relativistic approximation for simplicity of notation, instead of a more rigorous ${\cal O}_{NR}$. In addition, it must be recalled that the approximations must be understood on matrix elements. First, in this limit, since ${\cal O}$ is of order $c/v$, one can neglect the lower power: ${\cal O}^2 \ll {\cal O}^4$. In addition, for ground state mesons, one must note that the difference between the $\tilde{\rho^2},\tilde{\sigma^2}$ and the usual quantities ${\rho^2}, {\sigma^2}$ (eqns.(\ref{diffrho2},\ref{diffsigma2}) must be neglected here because they are subleading in $v/c$.

For instance, for one light quark (a ground state meson), one finds readily:
\begin{eqnarray} \label{curvatureNR1}
\sigma^2 \approx \tilde{\sigma^2}  \approx \frac{1}{3} m^4 (0|z^4|0) 
\end{eqnarray}.
This very simple NR expression for mesons (\ref{curvatureNR1}) has been presented and discussed in our papers (\cite{NRbounds,sigmaM}). It leads very straightforwardly to a sum rule involving $(\rho^2)^2$ and a sum on radial excitations:  
\begin{eqnarray}\label{SRcurvatureNR1}
\sigma^2 \approx \frac {3}{5} ((\rho^2)^2 +\sum_{n > 0} (\rho^{2 (n)})^2)
\end{eqnarray}
as well as to analogous sum rules for still higher derivatives.

For any number of light quarks, the corresponding expression for a $j=0$ state (baryon) is ;
\begin{eqnarray} \label{curvatureNR2}
\sigma^2=\tilde{\sigma^2} \approx\frac{1}{3}(0|{\cal O}^4|0) =\frac{1}{3}  (0|(\sum_k m_k z_k)^4|0) 
\end{eqnarray}

To obtain such a sum rule for any number of quarks, a general method is the following. It is useful to present it, because it can be generalised to exact BT outside of the NR approximation and to HQET (see next sections). One would like to relate the sum (without orbital excitations in intermediate states) $\sum_n |(j=0,n|({\cal O})^2|0)|^2$ to $(0|({\cal O})^4|0)$ (we write here parentheses for powers to avoid confusion with superindices).

This can be performed along the following lines. Let us define $\vec{{\cal O}}=\sum_i m_i \vec{r}_i$, whose ${\cal O}$ is the third component ${\cal O}={\cal O}^3$. Then $(j=0,n|({\cal O})^2|0)=\frac{1}{3} (j=0,n||\vec{{\cal O}}|^2 |0)$ with $|\vec{{\cal O}}|^2=({\cal O}^1)^2+({\cal O}^2)^2+({\cal O}^3)^2$. Moreover, $|\vec{{\cal O}}|^2$ has  transitions only to $j=0$ sates, therefore :
\begin {eqnarray}
\sum_n |(j=0,n|{\cal O}^2|0)|^2=\frac{1}{9}\sum_n |(j=0,n||\vec{{\cal O}}|^2|0)|^2=\frac{1}{9} (0||\vec{\cal O}|^2|\vec{\cal O}|^2|0)
\end{eqnarray}
. But the latter expression can be related to $(0|({\cal O})^4|0)$ by introducing a tensor:
\begin {eqnarray} \label{TNR}
T^{lmrs}=(0|{\cal O}^l{\cal O}^m {\cal O}^r {\cal O}^s|0)
\end{eqnarray}
One has : 
\begin {eqnarray}
(0|({\cal O})^4|0)=T^{3333}
\end{eqnarray}
while :
\begin {eqnarray}
(0||\vec{{\cal O}}|^2 |\vec{{\cal O}}|^2|0)=T^{llrr}
\end{eqnarray}.
This tensor is $SO(3)$ invariant since it is a matrix element over  $j=0$ states. And it is completely symmetric, because the components of $\vec{{\cal O}}_{NR}$ commute \footnote{Rigourously speaking, this means that the matrix elements of the commutator of two components of $\vec{{\cal O}}$ are of higher order in the $v/c$ expansion in internal velocities} . There is only one such tensor at each order, up to a constant : $T^{{lmrs}} \propto (\delta^{lm} \delta^{rs}+\delta^{lr} \delta^{ms}+\delta^{ls} \delta^{mr})$. The constant can be calculated from a trace:
\begin {eqnarray}
 T^{{lmrs}}&=&\frac{1}{15}~(\delta^{lm} \delta^{rs}+\delta^{lr} \delta^{ms}+\delta^{ls} \delta^{mr}) T^{llrr}
\end{eqnarray}
where :

One has then:
\begin {eqnarray}
\sigma^2&=&\frac{1}{3} T^{3333}=\frac{1}{15} T^{llrr}=\frac{1}{15}(0||\vec{{\cal O}}|^2 |\vec{{\cal O}}|^2|0)=\frac{9}{15} \sum_n |(0|{\cal O}^2|j=0,n)|^2 \nonumber\\&=&\frac{3}{5}((\rho^2)^2 +\sum_{n > 0} (\rho^{2 (n)})^2)  
\end{eqnarray}

In both HQET and exact BT (i.e. without NR approximation) cases, the new feature is that the various components of the tensor do not commute any longer, and this complicates somewhat the argument, although the general line is the same as above. The commutation laws are the same in BT and HQET : they are the ones of the Lorentz group.

\section{Developments common to BT approach and HQET} \label{HQET}

\subsection{Extension of the formulation to HQET}

In fact, nothing in the subsections \ref{exp},~\ref{higher} depends on the particular structure of ${\cal O}$ \footnote{in contrast, of course, with the results of subsection \ref{higherNR} concerning the NR limit. Also, in sections \ref{recall} to \ref{generalrho2}, the decomposition into ${\cal O}^{z,T}$ was used in certain places. The demonstration of the Bjorken sum rule does not require it, as shown again in \ref{higher}}. 
Then, it should be clear by now that all their results can be immediately extended from BT to general HQET Isgur-Wise functions by the simple substitution of the generator of an arbitrary Lorentz group representation instead of the particular one of BT:
\begin{eqnarray} \label{OK3}
{\cal O} \rightarrow {\cal K }^3
\end{eqnarray}
where index $3$ denotes the generator for the representation of a Lorentz transformation along the $z$-axis \footnote{The axis of Lorentz transformations are denoted by $x,y,z$ but those of the representation by $1,2,3$
to avoid confusion with the ordinary space upper indices like in ${\cal O}^z$, which denotes the $m=0$ part of ${\cal O}$. See footnote (\ref{indices}). Moreover, the representation of the Lorentz generator $K^i$ in the light quark space is written with calligraphic ${\cal K}^i$}.

And, indeed, in the formalism of Falk, as we have made explicit in \cite{LorentzI,LorentzII}, one can represent the generic Isgur-Wise function as :
\begin{eqnarray} \label{overlap}
\tilde{\xi}_{n \to n'}=(n'|U^\dagger(\Lambda') U(\Lambda)|n)=(n'|U({\Lambda'}^{-1}\Lambda)|n)
\end{eqnarray}
where $\Lambda,\Lambda'$ are Lorentz transformations allowing to pass from rest to velocities $v,v'$, say along $Oz$, and  $U(\Lambda),U(\Lambda')$  are the Lorentz representation matrices which allow to give these velocities  to the states $n,n'$. And obviously, if $v,v'$ are collinear, $U({\Lambda'}^{-1}\Lambda)$ may be cast in the exponential form with the help of the generator ${\cal K }^3$. Following the steps of subsection (\ref{exp})~(cf. eq. \ref{xinn'1}), we get now, with ${\cal O} \rightarrow {\cal K }^3$ :
\begin{eqnarray} \label{xiOK3}
\tilde{\xi}_{n \to n'}=(n'| e^{i~ \mathrm{argtanh}(\frac{v'-v}{1-v'v}){\cal K}^3} |n)
\end{eqnarray}

\subsubsection{Composition law and sum rules at finite $w$} 

Note that using the exponential expression (\ref {xinn'3vel}), one can formulate very simply the sum rules for finite velocities in matrix form :
\begin{eqnarray} \label{product}
\tilde{\xi}_{n \to n',v,v'}=\sum_{n"}  \tilde{\xi}_{n \to n",v,v"}  \tilde{\xi}_{n"\to n',v",v'}
\end{eqnarray}
This is exactly the expression of the HQET sum rules after factorisation of the heavy currents.
The simplest non trivial ones are with $n'=n=0,v'\neq v$:
\begin{eqnarray}  \label{product0}
\tilde{\xi}_{0 \to 0}(v,v')=\sum_{n"} \tilde{\xi}_{0 \to n"}(v,v")~\tilde{\xi}_{n" \to 0}(v",v').
\end{eqnarray}
which are the well-known "inclusive" sum rules of Bjorken, Isgur and Wise. 

The sum rule equation (\ref{product}) reflects the composition law of the representation of the subgroup. Indeed, it can be rewritten in matrix form, with the help of eq. (\ref{xinn'3vel}) as :
\begin{eqnarray} \label{matrixU}
U_{n',n}\left (\frac {v'-v}{1-v'v} \right )=\sum_{n"} U_{n',n"}\left (\frac {v'-v"}{1-v'v"}\right )~U_{n",n}\left (\frac {v"-v}{1-v"v}\right )~
\end{eqnarray}
where the argument of each $U$ is the relative velocity between the corresponding final and initial states.

From the above compact representations, one can deduce a variety of relations implied by HQET. One must warn that what will be obtained is not the full HQET results, but what results only from symmetry considerations in a broad sense (in older terms, Heavy quark symmetry), and not of course the fully dynamical part, which is a limiting form of QCD, neither the sum rules implying the binding energies of hadronic states (Voloshin sum rule and other "moments sum rules").

It must also be said that, by the present method, we do not obtain actually in HQET other results than the known ones. The advantages of the method, however, are real. They would reside first in simplifying the general presentation and the demonstration of known rules, as we now show on examples~-~mainly by extraction of the factors depending on velocities : this allows for the use of simpler, more direct algebraic methods, using directly $O(3)$ bases of states. The method is able to derive systematically a set of sum rules for derivatives, although the actual derivation implies an increasing complexity.
Moreover, as we shall develop in the last subsection, apart from the methodology, a main new fact resides in the direct transposition of HQET results to the BT quark models for any number of quarks.

\subsection{A selection rule : the transitions from a $j=0$ state} \label{selection}

One can first formulate selection rules based only on the angular momentum properties of the operator$({\cal K}^3)^p$ and the states.

As an example, it is shown very simply that transitions from a $j=0^+$ ground state to $j=0^-,2^-$ excited states, as well as any transition from a $j=0^+$ with change of naturalness, vanish at any $w$ in HQET as well as in the BT framework. This is of course valid for any heavy bilinear in the heavy quark mass limit. The selection rule was formulated first in the paper of Isgur, Wise and Youssefmir on baryons \cite{youssefmir}. See also section VI of Falk and Neubert \cite{neubert}.

Expanding the exponential in eq. (\ref{xinn'}) in power series, one sees that $\xi_{n\to n'}$ is a sum of $n',~n$ matrix elements of powers of ${\cal K}^3$ :
\begin{eqnarray} \label{powers}
\xi_{n \to n'}=\sum_{\nu} c_{\nu}(w) (n'|({\cal K}^3)^{\nu}|n)
\end{eqnarray}

The $c_{\nu}(w)$ contain increasing powers of $\sqrt{w-1}$. But we need not detail here their expression, coming from the expansion of the $argtanh$ factors. Having factorised the $w$ dependence in the coefficients $c_{\nu}(w)$, which are independent of the states, we have to discuss only the factors $(n'|({\cal K}^3)^{\nu}|n)$.

\subsubsection{$O(3)$ tensor description of angular momentum, parity and the theorem of conservation of naturalness}

Then, as to these factors, one can use purely algebraic  $O(3)$ arguments. 
Let us consider integer angular momentum of the light quarks. The angular momentum eigenstates of hadrons can be described by tensors of order $j$ or $j'$ of $SO(3)$.
$({\cal K}^3)^{\nu}$ is itself a component of a tensor of degree $\nu$, since ${\cal K}^3$ is the third component ${\cal K}^3$ of a 3-vector $\vec{{\cal K}}$ in full $SO(3)$ (with negative spatial parity)\footnote{One must be aware that ${\cal K}^3$ is an operator, and so is any component ${\cal K}^k$, but then the ${\cal K}^i,{\cal K}^k$ with $i \neq k$ do not commute, they obey the Lorentz generators algebra.}. 

The matrix elements $(n'|({\cal K}^3)^{\nu}|n)$ in eq. (\ref{powers}) representing the IW function are $SO(3)$ invariant under simultaneous rotations of the states and the operators because of space integration and spin summation. Let $j=0$, then, they can be formed only if one can saturate the indices of $(j'|$ by a number  $j'$ of those indices in the product $\prod_{m={1,...,\nu}} {\cal K}^{i_{m}}$, the $\nu-j'$ remaining ones in $\prod_{m={1,...,\nu}} {\cal K}^{i_{m}}$ being contracted with each other. This requires :
\begin{eqnarray} \label{jcons}
\nu =j'+2p~(p \geq 0)
\end{eqnarray}

Indeed, one requires at least $\nu_0=j'$ factors of ${\cal K}^i$ to saturate the indices of the final state. The additional " $2p$ " term is present because we may add  pairs of ${\cal K}^i$ with contracted indices in the product of the ${\cal K}$'s since such pairs are $SO(3)$ invariant, and, therefore, additional pairs of ${\cal K}^3, {\cal K}^3$ cannot be excluded.

On the other hand, let us consider space parity, acting on internal space variables. Let $\Pi,\Pi'$ be the parities of the initial and final states. $({\cal K}^3)^{\nu}$ has parity $(-1)^{\nu}$. Then, one must have:
\begin{eqnarray} \label{picons}
\Pi~\Pi'=(-1)^{\nu} 
\end{eqnarray}
Combining the eqns. (\ref{jcons},\ref{picons}), one obtains :
\begin{eqnarray} \label{naturalnesscons}
\Pi~\Pi'(-1)^{j'}=1
\end{eqnarray}
For $j=0$, $\Pi$ is also the naturalness for the initial state and $\Pi'(-1)^{j'}$ is the naturalness of the final state.
This means that if $j=0$ in the initial state, naturalness must be conserved in the heavy quark limit of BT, and in HQET as well, where it is well-known to be valid \cite{youssefmir, neubert}. In particular, $0^+$ cannot have transition to $0^-,2^-,...$. 

Let us stress that this is valid for any $w$ since the matrix elements $(n'|({\cal K}^3)^{\nu}|n)$ do not depend on it. Of course it is also valid for any current since the heavy quark current has been factorised, and its particular type does not appear.

\subsection{Sum rules for derivatives at $w=1$. Curvature sum rule} \label{curvatureHQET}

In analogy with subsection \ref{higher}, the derivatives are combinations of terms of the type $(n'|({\cal K}^3)^p|n)$\footnote{These powers can be regarded as $n!$ times the derivatives with respect to $\mathrm{argsinh}(\sqrt{w^2-1})$ and are elementarily related with those with respect to $w-1$, this implying additional terms of lower power.}. A set of sum rules for derivatives can be obtained just by inserting arbitrarily the closure relation of internal states inside the powers $p,...$ of ${\cal K}^3$: 
\begin{eqnarray} \label{decompO}
(n'|({\cal K}^3)^p|n)=\sum_{n"} (n'|({\cal K}^3)^{p_1}|n") (n"|({\cal K}^3)^{p_2}|n))
\end{eqnarray}
with $p_1+p_2=p$.
These sum rules can be obtained by simple algebra of the powers of ${\cal K}^3$, without involving the velocities. The most simple of them is the Bjorken sum rule written above, eq. (\ref{rho2}), which derives immediately from the expression of the first derivative $\tilde{\rho^2}=(0|{\cal O}^2|0)$ or more generally $(0|({\cal K}^3)^2|0)$, and which we have discussed in detail in the preceding sections.

In the case of the Bjorken sum rule, the derivative $\rho^2$ is related to transitions with a different angular momentum, $j=1$. For higher, even order, derivatives, it is possible to relate them to lower derivatives for states with the same angular momentum, including radial excitations.

\subsubsection{Sum rule for curvature and radial excitations} 

As already recalled, an important sum rule of this type is the one demonstrated for HQET in the paper \cite{sigmaB}) for the curvature of $j=0$ baryons :
\begin{eqnarray} \label{curvature}
\sigma^2=\frac {3}{5}  \left ( (\rho^2)^2+ \sum_{n \geq 1}|\xi^{'(n)}(w=1)|^2 + \rho^2 \right )
\end{eqnarray}
We can redemonstrate it by the present operator method in a more direct manner. However, this demonstration is still more complicated than the one for Bjorken, in particular because one must now recourse to other components of $\vec{{\cal K}}$, and then, unlike in the NR development of subsection \ref{higherNR}, one cannot use commutativity.

The initial steps are the same as the ones in subsection \ref{higherNR}. The first derivatives for an elastic Isgur-Wise function or transition to a radially excited state correspond to $p=2$. We would like to calculate a sum on such derivatives, excluding orbital excitations:
\begin{eqnarray} \label{sum0}
{ \sum}_{n}(0|({\cal K}^3)^2|n,j=0)(n,j=0|({\cal K}^3)^{2}|0)
\end{eqnarray}. 

We cannot directly apply eq. (\ref{decompO}) because we have restricted the sum to $j=0$ states. However, our ${\cal K}^3$ is the 3 component of the vectorial operator $\vec{{\cal K}}$, we can again use closure over $j=0$ states after having related the matrix elements in eq. (\ref{sum0}) to a spherically symmetric operator, who has transitions only to $j=0$:
\begin{eqnarray}
(n,j=0|({\cal K}^3)^2|0)= \frac{1}{3} (n,j=0||\vec{{\cal K}}|^2)|0)
\end{eqnarray}
whence using closure on all the states $n'$, one gets: 
\begin{eqnarray} \label{sum0prime}
\sum_{n}(0|({\cal K}^3)^2|n,j=0)(n,j=0|({\cal K}^3)^2|0) \nonumber\\
=\frac{1}{9} \sum_{n'} (0|(\vec{{\cal K}})^{2}|n')(n'|(\vec{{\cal K}})^{2}|0)=\frac{1}{9} (0|(\vec{{\cal K}})^{2}(\vec{{\cal K}})^{2}|0)
\end{eqnarray}
To proceed further, we use again the fact that the last matrix element in eq. (\ref{sum0prime}) may be related  to the analogous fourth power matrix element $(0|({\cal K}^3)^4|0)$ controlling $\sigma^2$. Both are expressible through components of the same invariant $O(3)$ tensor:
\begin{eqnarray}
T^{lmrs}=(0|{\cal K}^{l}...{\cal K}^{s}|n,j=0)
\end{eqnarray} 
with $l,m,r,s=1,2,3$, generalising eq.(\ref{TNR}). $(0|({\cal K}^3)^4|0)$ is the $T^{3333}$ component, while  $(0|(\vec{{\cal K}})^{2}(\vec{{\cal K}})^{2}|0)$ is a trace over the same tensor :
\begin{eqnarray}
(0|(\vec{{\cal K}})^{2}(\vec{{\cal K}})^{2}|0)=T^{llrr}
\end{eqnarray} 

This tensor is again $O(3)$ invariant because the matrix elements are taken on $j=0$ states. The two quantities $T^{3333}$, $T^{llrr}$, one controlling the sum, and the other the curvature, are therefore connected as in the previous section. 
Now, the generators ${\cal K}^i$ no longer commute, and then the tensor $T^{i_1 i_2....i_{2p"}}$ is no longer fully symmetric. Then, there are several distinct invariant tensors. This complicates the deduction of the sum rule. The calculation now implies the use of the Lorentz Lie algebra (see eqn.(\ref{KKJ}) below). Happily, the commutators of two components of $\vec{{\cal K}}$ are just angular momentum operators, therefore it remains tractable, and it appears that the r.h.s. of eq.(\ref{sum0prime}) can be entirely expressed in terms of matrix elements over the ground state of powers $\leq 4$ of ${\cal K}^3$, i.e. curvature and slope of the Isgur-Wise function.
These commutators add lower powers of ${\cal K}^3$ to the former expressions obtained with the non-relativistic assumption of commutability because they correspond to relativistic corrections in a $v/c$ expansion of matrix elements (remember that ${\cal O}$ is $O(c/v)$, therefore lower powers of ${\cal O}$ are suppressed by powers of $v/c$).

\subsubsection{Reduction of the auxiliary tensor by Lorentz group commutation relations}



T is an even invariant tensor under $O(3)$, therefore it can be decomposed into :
\begin{eqnarray}
T^{i_1 i_2 i_3 i_4}=A \delta^{i_1 i_2} \delta^{i_3 i_4}+B\delta^{i_1 i_3} \delta^{i_2 i_4} +C \delta^{i_1 i_4} \delta^{i_2 i_3}
\end{eqnarray}
One can determine the coefficients $A,B,C$ in function of the 3 traces of $T$,
$T_i$, which are more directly connected to the concerned quantities. Whence three equations :
\begin{eqnarray}
T_1=T^{i_1 i_1 i_2 i_2}=9 A+3 B+3 C \label{T1}\\
T_2=T^{i_1 i_2 i_1 i_2}=3 A+9 B+3 C \label{T2}\\
T_3=T^{i_1 i_2 i_2 i_1}=3 A+3 B+9 C \label{T3}
\end{eqnarray}
$T_1$ is $T^{i_1 i_1 i_2 i_2}=(j=0|\vec{{\cal K}}^2 \vec{{\cal K}}^2|j=0)$. $T^{3 3 3 3}=A+B+C=\frac{1}{15}(T_1+T_2+T_3)$.

The problem is therefore solved if we can relate $T_{2,3}$ to $T_1$. This can be obtained by use of the commutation relations of Lorentz group generators \footnote{One can find a rather complete survey of properties of the Lorentz groups, generators, Casimir operators, etc... in the book of Lomont, reference \cite{lomont}}. First $T^{i_1 i_2 i_1 i_2}=T^{i_1 i_2 i_2 i_1}$, i.e. $T_2=T_3$; indeed one can exchange the last two indices because 
\begin{eqnarray} \label{KKJ}
[{\cal K}^{i_2},{\cal K}^{i_1}]=-i \epsilon^{i_2 i_1 i_5} {\cal J}^{i_5}
\end{eqnarray}
and ${\cal J}^{i_5}|j=0)=0$. Therefore, $B=C$ (compare eqns. (\ref{T2},\ref{T3})) . 

One may now eliminate $T_3$ to remain with only one independent quantity $T_1$. The dissymmetry
between $T_1$ and $T_3$ is due to the non commutativity of ${\cal K}^{i_2},{\cal K}^{i_1}$. We exploit the commutativity of the Casimir $\vec{{\cal K}}^2-\vec{{\cal J}}^2$ with all the Lorentz generators :
\begin{eqnarray} \label{T3T1}
T_3-T_1=(T^{i_1 i_2 i_2 i_1}-T^{i_1 i_1 i_2 i_2})=\nonumber\\(j=0|{\cal K}^{i_1}[\vec{{\cal K}}^2,{\cal K}^{i_1}]^{-}|j=0)=\nonumber\\
(j=0|{\cal K}^{i_1}[\vec{{\cal J}}^2,{\cal K}^{i_1}]_{-}|j=0)=\nonumber\\
2~(j=0|{\cal K}^{i_1}{\cal K}^{i_1}|j=0)=6 \rho^2
\end{eqnarray}
The last but one equality comes from :
\begin{eqnarray}
&\vec{{\cal J}}^2|j=0)&=0,\nonumber\\
&\vec{{\cal J}}^2 ({\cal K}^{i_1}|j=0)&=\vec{{\cal J}}^2|j=1)=1(1+1)|j=1)=2|j=1) .
\end{eqnarray}

Finally, one finds, solving for $A,B,C$ in terms of $T_1$:
\begin{eqnarray} \label{finalT4}
T^{i_1 i_2 i_3 i_4}=\frac{1}{15} T_1 ( \delta^{i_1 i_2} \delta^{i_3 i_4}+\delta^{i_1 i_3} \delta^{i_2 i_4} + \delta^{i_1 i_4} \delta^{i_2 i_3}) 
+\nonumber \\
\frac {3} {15} \rho^2 (-\delta^{i_1 i_2} \delta^{i_3 i_4}+\frac{3}{2} (\delta^{i_1 i_3} \delta^{i_2 i_4} + \delta^{i_1 i_4} \delta^{i_2 i_3}))
\end{eqnarray}

\subsubsection{The resulting HQET sum rules}
{\it Main sum rule}

From eqn. (\ref{finalT4}), all the relevant quantities can now be expressed in terms of $T_1$ (and of course $\rho^2$). First :
\begin{eqnarray}
(j=0|({\cal K}^3)^4|j=0)=\frac{1}{5} T_1 +\frac{4}{5}\rho^2 \\
\sum_{n} |(n,j'=0|{\cal K}^3)^2|j=0)|^2= \frac{1}{9}(j=0|\vec{{\cal K}}^2 \vec{{\cal K}}^2|j=0)= \frac{1}{9}T_1
\end{eqnarray}
This yields immediately by elimination of $T_1$:
\begin{eqnarray}
(j=0|({\cal K}^3)^4|j=0)= \frac {9}{5}\sum_{n} |(n,j'=0|({\cal K}^3)^2|j=0)|^2+\frac{4}{5}\rho^2
\end{eqnarray}
\begin{eqnarray}
\sigma^2=\frac {(j=0|({\cal K}^3)^4|j=0)+\rho^2}{3}= \nonumber \\
\frac {3}{5} \left (\sum_{n} |(n,j'=0|({\cal K}^3)^2|j=0)|^2+\rho^2 \right )
\end{eqnarray}

\noindent {\it Relation between $j'=0$ and $j'=2$ transitions, and the second sum rule.}

One gets also immediately a relation between the $j'=2$ and $j'=0$ contributions :
\begin{eqnarray}
\sum_{n}|(n, j'=2|({\cal K}^3)^2|j=0)|^2=\frac{4}{5} \left (\sum_{n} |(n,j'=0|({\cal K}^3)^2|j=0)|^2+\rho^2 \right )
\end{eqnarray}
and finally :
\begin{eqnarray}
\sigma=\frac{4}{3} \sum_{n}|(n, j'=2|({\cal K}^3)^2|j=0)|^2
\end{eqnarray}
which is found, written with another notation, in the paper \cite{sigmaB}, just before the main rule (equation (56) of this paper).

\subsection{Automatic validity of the above HQET selection rules and sum rules for BT quark models}
\label{automatic}



Presently, the main advantage of the method, in our opinion, concerns the demonstration of the validity of a large class of HQET results in the BT approach to quark models. Indeed, conversely, all the properties and relations of HQET which derive from the expressions of the former subsection or more generally of the representation eqn.(\ref{xiOK3}), are automatically valid for the Bakamjian-Thomas models as well, since one has just to particularise by the inverse substitution 
\begin{eqnarray}\label{HQETtoBT}
{\cal K}^3 \to {\cal O}=\sum_k {\cal O}_{k}
\end{eqnarray}
where each ${\cal O}_{k}$ is given by the expression (\ref{calO}) applied to each light quark $i$. Anyway, the reference to this particular expression is purely formal, since in these rules we have not recoursed to any specific feature of ${\cal K}^3$~~\footnote{\label{K3} This arbitrariness must be understood with the restriction that ${\cal K}^3$ must belong to a full set of operators satisfying the Lorentz group Lie algebra, since covariance is implied in HQET. And then ${\cal O}$ should share the same property. But as already stressed, ${\cal O}$, which is one among the set of generators of Lorentz transformations for free particles, satisfies automatically this condition, and this amounts to a covariance property for BT also in the heavy quark limit, as discussed in the next section.}. 

This automatic implementation in the BT approach applies for the HQET sum rules and selection rules of the previous subsections. It is then very useful because it would be very difficult to demonstrate them for systems with several light quarks, for instance for baryons, by the methods that we have previously used in BT, i.e. by obtaining manifestly covariant expressions for the Isgur-Wise functions of each particular state. It must be stressed that these results seem also to be an important advantage of the BT approach to quark models over other quark model approaches.

These arguments do not apply however to the so-called "moment" sum rules. Moment sum rules must be considered as related to non trivial powers in $1/m_Q$ expansion, as is the case in the OPE demonstration. Only those sum rules corresponding to the strict heavy quark limit, with the expressions like (\ref{xiOK3}), apply to BT. The moment sum rules involve not only the Isgur-Wise functions, but the binding energies and the binding energies are not present in BT, which also becomes not covariant at finite mass. 

\section{Covariance of the BT approach in the heavy quark limit} \label{covariance}
 
Let us recall that full covariance of the BT approach in the heavy quark limit is clearly exhibited for particular cases (ground state meson I-W function, transitions to $1/2,3/2,L=1$ mesons, and now baryons in a next paper) in the manifestly covariant formulation (see the references at the beginning of the present paper), obtained by translating the two dimensional spinor description into a Dirac-like formulation- with four-dimensional polarisation tensors and spinors. We now show that the present formulation allows to show that this holds for states with arbitrary angular momenta and number of light quarks.

\subsection{Manifest covariance for collinear transformations}
But let us first stress that,
although expressed in unfamiliar terms, eq. (\ref{xinn'}), as well as eq. (\ref{matrixU}) check the covariance of the formalism in a manifest form, albeit limited to the collinear group of Lorentz transformations. 

It must be indeed underlined that in the formalism the states  $|n),|n')$ do not depend on hadron motion. Therefore, any dependence on frame may come only from the exponential operator inserted between the two states of the matrix element, and we find that in fact {\it this operator depends only on the invariant $w$}. Therefore the matrix element itself is invariant. But invariance is exactly what is expected from covariance under the collinear subgroup : indeed spin states should be transformed through Wigner rotations, but these are trivial for a Lorentz transformation along the spin projection axis. 

\subsection{Full covariance}

But finally we attract attention to the fact that full covariance of the BT formalism in the heavy quark limit for arbitrary states is established by appealing to the general demonstration of the Lorentz covariance of overlaps given in our paper \cite{LorentzI}, sections 2.1 and 2.2. 

One notes that this demonstration rests entirely on an expression for HQET overlaps
identical to the one given above in the present paper, eqn. (\ref{overlap}): 
\begin{eqnarray} \label{overlap2}
\tilde{\xi}=(n'|U^\dagger(\Lambda') U(\Lambda)|n)
\end{eqnarray}
$U(\Lambda^{,'})$ being the Lorentz transformations for the velocities $v,v'$.
It is demonstrated that such an expression ensures covariance.

Note however that, in this demonstration, $v,v'$ should not be assumed to be collinear, unlike what we assumed up to now in the present paper. But one easily sees that the expression (\ref{overlap2}) holds for the overlaps of BT quark models also when the velocities are not collinear. Indeed, we can apply the arguments and calculations of sections \ref{first} and \ref{IW} which concern only one state separately, to a velocity along any axis other the initial $Oz$, and this independently for the initial and final state, therefore even for non collinear velocities. Denoting the unitary vector for the direction of $v$ as $\hat{v}$, one defines for a general direction $\hat{v}$:
\begin{eqnarray}
{\cal O}^{\hat{v}}=\vec{{\cal O}}.\hat{v}
\end{eqnarray}
which means that in the expression of ${\cal O}$, $Oz$ is everywhere replaced by the generic direction $\hat{v}$: $z \to \vec{r}.\hat{v}$, $(\vec{\sigma} \times \vec{p})^z \to (\vec{\sigma} \times \vec{p})^{\hat{v}}$. The operator realising the generic $U(\Lambda)$ is then :
\begin{eqnarray}
U(\vec{v})=e^{-i~\mathrm{argtanh}(v) \vec{{\cal O}}.\hat{v} }
\end{eqnarray}
instead of $U(v)$ of eq. (\ref{U}) and the analogous for $v'$. $\tilde{\xi}$ has now the form :

 \begin{eqnarray}
\tilde{\xi}_{n \to n'}(\vec{v},\vec{v'})=(n'|U^{\dagger}(\vec{v'}) U(\vec{v})|n)=(n'|e^{i~ \mathrm{argtanh}(v') \vec{\cal O}.\hat{v'} } e^{-i~ \mathrm{argtanh}(v) \vec{\cal O}.\hat{v} }|n)
\end{eqnarray}
with $\Lambda \to U(\vec{v})$ realising the representation of Lorentz $U(\Lambda)$ for any direction of $\vec{v}$ , and $\tilde{\xi}$ having the form (\ref{overlap2}) required for full covariance.

As to the full current, it is then obtained by multiplying the overlap and the heavy quark current, which is covariant by itself in a trivial manner. Whence full covariance follows for all the heavy current matrix elements.



\section{Conclusion} 
Starting from an analysis of the Bakamjian-Thomas quark models in the heavy quark limit, we have shown that one can describe the Isgur-Wise functions of hadron transitions in the heavy quark limit by considering only the light quark system (of course submitted to the field of a static source). This result corresponds to the old general analysis by Falk in HQET, with his central idea that the dynamical part of the Isgur-Wise functions are just overlaps of states, which are themselves of course obtained by Lorentz boosts. 

As a second step also suggested by the analysis of Bakamjian-Thomas quark models, 
we propose to reduce these overlaps to matrix elements of states at rest by the use of the infinitesimal Lorentz group generators.

The advantages and limitations of the present method for BT and HQET consist in the following points.

\begin{enumerate}
\item
The simplicity and physical intuitiveness of the expressions for the slopes or higher derivatives of Isgur-Wise functions, and the formal, exponential expression of Isgur-Wise functions. These are extensions of the corresponding 
non relativistic expressions for form factors:
\begin{eqnarray}
(n'|e^{i k z}|n)=(n'|e^{i m (v'-v ) z}|n)
\end{eqnarray} 
and their derivatives (which are powers of $z$). One needs only the intuitive substitutions:
\begin{eqnarray} \label{NR-Rel}
\sum_k m_k z_k \to {\cal O}_{BT}~~\mathrm{or}~~{\cal K}^3
\end{eqnarray} 
for the dipole operator, and : 
\begin{eqnarray} 
v'-v \to \mathrm{argtanh}\left(\frac{v'-v}{1-v'v}\right)
\end{eqnarray} 
for the relative velocity factor.
 
The simplicity is allowed furthermore by the choice of the collinear frames.

The case of Bakamjian-Thomas is especially simple because of the transparent structure of ${\cal O}_{BT}$ as the generator of Lorentz transformation for free light quarks (see further comments in item \ref{itemBT}). 

\item A description of states which involves only their $O(3)$ rest frame structure, without reference to the motion. The motion appears only through universal functions of $w$. 
\item
Generality  as regards possible angular momentum of states and the number of quarks, all of them being treated in the same way and with similar expressions, as the standard series of representations of $O(3)$.

One avoids the complications of the Rarita-Schwinger formalism. 

\item 
Rules of heavy quark symmetry demonstrated in a simple purely algebraic way, just by handling powers
of the components of $\vec{{\cal O}}$ or $\vec{{\cal K}}$, the prototype being the Bjorken sum rule 

\item The Isgur-Wise functions are decomposed into a power series with coefficients factorised into universal factors, independent of the type of state, depending on $w$ only (in fact powers of $\mathrm{argsinh}(\sqrt{w^2-1})$), times matrix elements of the corresponding powers of ${\cal O}$. Using this expansion, one may obtain rules valid for arbitrary $w$, the example being the selection rule on naturalness, see subsubsection \ref{selection}.

\item  
An exceptionnally easy transposition of heavy quark symmetry rules (HQS) from BT to HQET or the reverse, BT at $m_Q=\infty$ appearing as a particular case of HQET with the only specification of the Lorentz generator as the free particle representation generator. The HQET sum rules of the Bjorken-Uraltsev type \footnote{By this, we denote the sum rules corresponding to the lowest order for an OPE moment expansion, which do not implies power of excitation energy $\Delta E$ } (e.g. Bjorken or curvature sum rules) are then at once valid for BT without reference to a particular number of light quarks. This is an important result for several light quarks ($\geq 2$) because they would be difficult to demonstrate by our standard method using manifestly covariant expressions of Isgur-Wise functions for the BT quark models. 

Conversely, rules found in BT whose existence depend only on the existence of an operator
${\cal O}$ without reference to a specific Lorentz group representation can be immediately transposed to HQET. 

Of course, the specific advantage of the BT quark models is that, beyond these general sum rules, they provide quantitative estimates.

\item \label{itemBT} Although one main interest of the formulation is its generality, which allows to establish a very direct connection with general HQET, there are also useful consequences coming from the specific structure of the operator ${\cal O}$ corresponding to the BT models, namely the generator of Lorentz tranformation along $z$ for free particles. Let us recall this structure. For one light quark, one has (eqn. \ref{calO}):
\begin{eqnarray}  
{\cal O}=-\frac{z p^0+p^0 z}{2}+\frac{1}{2} \frac {(\vec{\sigma} \times \vec{p})^z} {p^0+m},
\end{eqnarray}
to be summed over light quarks when several are present.
This formula displays intuitively the way relativistic effects are intervening in formula (\ref{NR-Rel}), starting from the NR expression $\sum_k m_k z_k$ :

$\alpha$) the first term corresponds to the NR expression $m~z$, with the mass factors $m$  substituted by free quark energies $p^0$.
The analysis shows that this substitution corresponds to the well-known effect of the spatial Lorentz transformation $z'=\frac {z-v t} {\sqrt{1-(v/c)^2}}$, as opposed to Galilean transformation
$z'=z-v t$. 

$\beta$) the second term corresponds to the Lorentz transformation of bidimensional spin.

In addition to the generic $j=1$ property, this operator has a very specific
$O(3)_{space} \times SO(3)_{spin} $ structure, which leads to obvious additional selection rules, and to  structurally different type of contributions. 

There may be also sum rules depending on this specific structure of BT, and therefore, which are specific to quark models. An example of such sum rules has been given in the paper : the sum rule similar to the Bjorken sum rule, except that the intermediate $j=1$ states are restricted to $j=1,s=1$ (see eqn. (\ref{wignerCov})). Of course such specific sum rules are not expected to be hold in exact HQET, and also eqn. (\ref{wignerCov}) involves the explicit knowledge of the ground state wave function : the sum is not equal to an observable.

\item The first merit is thus algebraic : it corresponds to the use of the $O(3)$ algebra, especially the tensorial algebra, or in certain cases the Lie algebra of the Lorentz group. 

There is also an important qualitative advantage in the possibility of displaying in a straightforward way the NR expansion for internal velocities, which gives a very useful intuition of the origin and effect of various terms: the powers of ${\cal O}$ are powers of $c/v$, and, inside ${\cal O}$, the Wigner rotation term is in $v/c$ compared to the main one which is $c/v$ ; in both terms, one gets kinetic energy correction to the mass factors of relative order $v^2/c^2$... 

But the approach could be also useful for certain quantitative calculations. One can say that quantitatively, the approach is suited for calculations near  $w-1$, i.e. requiring a low number of powers of ${\cal K}^3$ (slopes, curvature), and especially advantageous for large number of light quarks (two and more), where, on the other hand, their expressions in the manifestly covariant approach becomes very complex...

The formulation emphasizes the initial system of coordinates $\vec{r_2},\vec{r_3},...$ rather than standard systems of relative coordinates like $\vec{\lambda}, \vec{\rho} $, i.e., it favours the relative coordinate of each light quark with respect to the static heavy quark.

On the other hand, at arbitrary $w$, the exponential expression is only a compact representation of a power series. Indeed, high powers require indefinitely high powers of derivatives, which are not numerically tractable.
Here, the manifestly covariant formulation, which we have developped since a long time, and which allows to express Isgur-Wise functions at any $w$ as simple integrals, is more suited. 

\item  As applications of this new formulation, we have only presented the ones related to "baryons", in the sense of states with integer $j=0,j=1$, but which may have any even number of light quarks. 

One would have later to present the ones relative to mesons ($j=1/2,3/2$), which have been presented shortly in a roughly similar language in the paper \cite{exact}. One should for instance retrieve the Uraltsev sum rule\footnote{In fact,this sum rule has already been demonstrated in BT for the case of ordinary, $Q \bar{q}$ mesons in both the manifestly covariant and the present methods \cite{nousUraltsev}}.
It may seem paradoxical that they are somewhat more complicated to handle than the baryons. This is simply because what is relevant in such mainly algebraic considerations is the angular momentum complication, not the number of quarks. Practical calculations with concrete wave functions is another question.

\end{enumerate}

\end{document}